# Spectral synthesis provides 2-D videos on a 1-D screen with 360°-visibility and mirror-immunity

Sascha Grusche

*Physikdidaktik, Fakultät 2 - Pädagogische Hochschule Weingarten, Kirchplatz 2, 88250 Weingarten, Germany*
*Corresponding author:* saschagrusche@gmail.com

Spatial-light-modulator (SLM)-based tunable sources have complex setups. A simpler setup, comprising an SLM-projector and a dispersive element, synthesizes light as effectively, based on a Superposition of Newtonian Spectra (SNS). As a generalization of SNS, two-dimensional (2-D) grayscale videos are spectrally encoded on a one-dimensional (1-D), translucent screen, and viewed through another dispersive element. This Projected-Image Circumlineascopy (PICS) produces semitransparent, rainbow-coloured, virtual 2-D videos that face every viewer anywhere around the 1-D screen. They are invariant under reflection of the 1-D screen in mirrors parallel to it. SNS bandwidth and PICS image geometry are calculated using geometric optics and Dispersion Diagrams.

## 1. Introduction

Light sources whose spectrum can be adjusted for specific wavelengths at specific intensities are needed for applications ranging from microscopy and endoscopy [1], to colorimetry and color imaging [2-4], to stage lighting [5] and hyperspectral imaging [6-8]. Such tunable light sources are typically based on light emitting diodes (LED)[9] or spatial light modulators (SLM) to generate light via spectral synthesis.

SLM-based tunable light sources synthesize light usually by subtracting multiple parts from a single spectrum: A white slit image, mostly from a xenon [2,4] or mercury arc lamp, is dispersed by prisms [2,5] or gratings [4,5,10] onto a digital mirror device (DMD) or liquid crystal device (LCD) [2] panel, which masks out parts of the spectrum. Each column of the SLM represents a specific peak wavelength, whose intensity is regulated either by the number of 'on' pixels in that column [2,10], or through pulse width modulation [5,11]. Thus, a spectrum, or rows of spectra, are synthesized. They are combined into a projected line [11], on a diffusing plate [2], in an integrating cavity [5], in an optical fiber, or in a liquid light guide [4]. For calibration and further use of the light engine, the spectral power distribution (SPD) of the light is measured with a camera [4] or a spectroradiometer [2,12]. An optional feedback loop optimizes the spectral output [2,4,12]. Essentially, such SLM-based light engines are conversions of a spectroscopic setup.

This conventional approach to spectral synthesis is intuitive and effective. Conversely, there are two drawbacks: A) A complex apparatus is needed for three major steps: 1) dispersion of white light, 2) SLM-based modification, 3) recombination. B) Because the SLM is used to encode wavelengths as two-dimensional (2-D) patterns, its potential to encode 2-D images as wavelengths [13] has not been realized with such light engines.

Correspondingly, I propose an SLM-based light engine that has two advantages: A) The setup is simpler, yet as effective. It synthesizes light by combining single parts from multiple spectra. This Superposition of Newtonian Spectra (SNS) takes only two steps: 1) SLM-based modification of white light, 2) dispersion. B) With the light engine, one can watch 2-D, mirror-immune videos from all around a 1-D projection screen, in a viewing method called Projected-Image Circumlineascopy (PICS).

In Section 2, I introduce SNS spectral synthesis. Building on SNS, I present PICS. Section 3 describes a setup for both SNS

and PICS. Section 4 contains the experimental results. In Section 5, I compare SNS to other methods of spectral synthesis, and discuss PICS image properties. Finally, I conceptually unify spectral synthesis with spectral encoding. Section 6 summarizes the concepts and findings, and suggests future research.

## 2. From SNS essentials to PICS pictures

### A. Spectral Synthesis based on a Superposition of Newtonian Spectra (SNS)

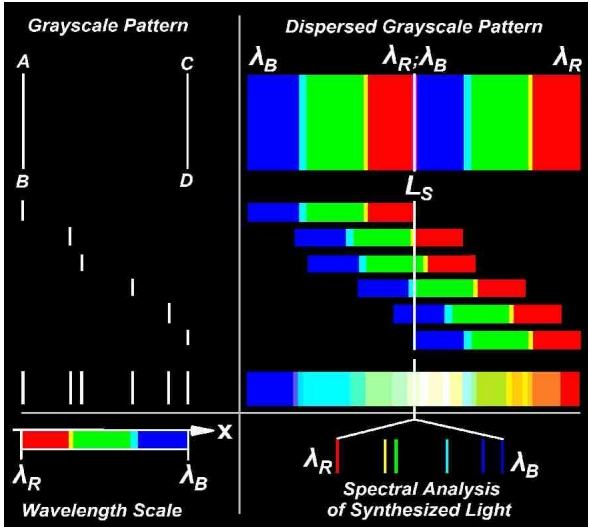

Fig. 1. In SNS, a grayscale pattern is dispersed to synthesize light at $L_S$.

*1. The essence of SNS*

Light of a desired wavelength composition can be synthesized by superposing (at a linear locus $L_S$) different color stripes from multiple Newtonian spectra (see Fig. 1). We obtain a single Newtonian spectrum by projecting a white slit image from a broadband source through a dispersive element (e.g. a prism or grating) [14]. Different color stripes in the spectrum represent different peak wavelengths, from $\lambda_B$ to $\lambda_R$. For multiple Newtonian spectra, we project multiple white lines through the same dispersive element for synthesis ($DE_S$).

For spectral synthesis, consider the geometry of spectra. Suppose we project a white line through a $DE_S$, whose distance from the projector is $d_P$, as shown in Fig. 2(a). Then, the rays for $\lambda_B$ and $\lambda_R$ emerge from $DE_S$ under a dispersion angle $\delta_S$. (Consider $\delta_S$ positive if the ray for $\lambda_B$ can be made to coincide with the ray for $\lambda_R$ by a clockwise rotation of less than 90° about the vertex.) At a distance $d_S$ from $DE_S$, the rays for $\lambda_B$ and $\lambda_R$ create two monochromatic images whose mutual displacement, the so-called dispersive displacement for synthesis, is

$$s = 2\tan(0.5\delta_S)d_S. \quad (1)$$

(If a ray passes a series of dispersive elements, we apply Eq. (1) sequentially.)

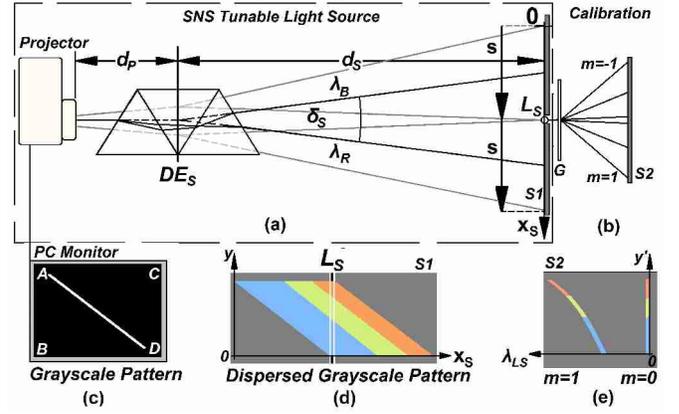

Fig. 2. Setup and wavelength calibration for SNS. (a)-(b) Top view. (b)-(d) Line AD is dispersed to obtain a calibration spectrum at $L_S$.

A white line (in the $y$-direction) of width $\Delta x$ at a projection distance $D_P = d_P + d_S$ is dispersed by $DE_S$ (in the $x_S$-direction) into a Newtonian spectrum of width

$$w_S = |s| + \Delta x. \quad (2)$$

The linear locus of spectral synthesis $L_S$ is the place where two Newtonian spectra of lines AB and CD kiss each other to synthesize light from the ends of the spectrum (see Fig. 1). The kissing condition is that the distance between lines AB and CD equals the dispersive spread [15], i.e. the absolute value of the dispersive displacement for synthesis:

$$\overline{AC} = \overline{BD} = |s|. \quad (3)$$

Under this condition, the spectra of lines AB and CD barely overlap in a narrow zone of width $\Delta x$, supplying peak wavelengths $\lambda_B$ and $\lambda_R$ to $L_S$ (in the $y$-direction). Then, intermediate spectra from lines between AB and CD contribute intermediate wavelengths to $L_S$. With lines AB and CD in place, we insert a grayscale pattern that resembles the desired slit spectrum, as in Fig. 1.

*2. Calibrating the grayscale pattern*

Having established lines AB and CD, we use a white line AD for wavelength calibration, and a white rectangle ABCD to calibrate the intensity of the grayscale pattern.

For the wavelength scale, we assign to each pattern coordinate $x$ the peak wavelength $\lambda_{LS}(x)$ obtained at $L_S$:

$$x = x(\lambda_{LS}) \quad (4)$$

Projected through $DE_S$, line AD ($y = kx$; $k$...any constant, see Fig. 2(c)) yields a calibration spectrum at $L_S$, as in Fig. 2(d). Its wavelengths $\lambda_{LS} = \lambda_{LS}(y)$ imply $\lambda_{LS} = \lambda_{LS}(x)$.

For the intensity $I(x)$ of the grayscale pattern, we relate the desired spectral intensity distribution $I_A(\lambda_{LS})$ to the spectrum of synthesized white light $I_{A,full} = I_{A,full}(\lambda_{LS})$:

$$I(x) = \mu \frac{I_A(\lambda_{LS})}{I_{A,full}(\lambda_{LS})} \quad (5)$$

The weighting function $\mu$ holds $I(x)$ below or at the intensity of white light, $I_{max}$, and compensates for the fact that each pixel contributes a certain range of wavelengths. Projected through $DE_S$, rectangle ABCD of intensity $I(x) = I_{max}$ yields $I_{A,full}$ at $L_S$.

## 3. Calculating SNS spectral bandwidth

With wavelengths $\lambda_{LS} = \lambda_{LS}(x)$, the bandwidths $\Delta\lambda_{LS}(x)$ follow from Dispersion Diagrams [15], as in Fig. 3.

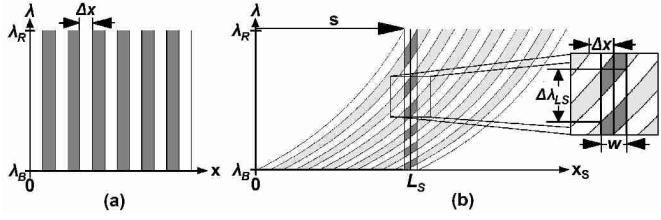

Fig. 3. Deriving SNS spectral bandwidth from Dispersion Diagrams. Dispersion vector **s** translates the pixel resolution of the grayscale pattern into the spectral resolution of the synthesized light at LS.(a) Wavelength distribution of alternating white and black pixels. 100% spectral intensity is symbolized by white areas, 0% by dark gray areas.(b) Wavelength distribution of the dispersed pixels (irrelevant parts at lower contrast).

To derive the wavelength distribution of the synthesized light, we shear that of the grayscale pattern from Fig. 3(a) according to dispersion vector $\mathbf{s} = s\,\hat{x}_S$; unit vector $\hat{x}_S$ pointing to the right on S1, see Fig. 2(a). If we assume equal dispersion for all coordinates x, the curvature of the sheared wavelength distribution in Fig. 3(b) mirrors the curve for $\lambda_{LS} = \lambda_{LS}(x)$.

Suppose the grayscale pattern has $N$ pixel columns. If a projected grayscale pixel is wider than the synthesized line of light ($\Delta x = w_S/N \geq w$), adjacent pixel columns from Fig. 3(a) translate into adjacent spectral lines in Fig. 3(b) whose mutual shift is one FWHM. Thus, a pixel between $x$ and $x+\Delta x$ yields a spectral line with peak wavelength $\lambda_{LS} = \lambda_{LS}(x)$ and bandwidth

$$\Delta\lambda_{LS}(x) = \left|\lambda_{LS}\left(x-\frac{\Delta x}{2}\right) - \lambda_{LS}\left(x+\frac{\Delta x}{2}\right)\right|\left(1+\frac{w}{\Delta x}\right) \quad (6)$$

$\lambda_{LS}'(x)$ being roughly constant across $w$.

If, to simplify, we assume linear dispersion

$$\lambda_{LS}(x) = \lambda_R - \left|\frac{(\lambda_R - \lambda_B)}{s}\cdot x\right|, \quad (7)$$

then, each pixel yields a spectral bandwidth

$$\Delta\lambda_{LS}(x) = \frac{|\lambda_R - \lambda_B|}{(N-1)}\left(1+\frac{w}{\Delta x}\right). \quad (8)$$

### B. SNS for Projected-Image Circumlineascopy (PICS)

With SNS, we obtain a spectral intensity distribution $I_A(\lambda_{LS}, y)$ at $L_S$ by projecting a grayscale pattern of intensity $I = I(x,y)$ through $DE_S$. In plain words, the slit spectrum of the synthesized line of light is *always* a rainbow-colored version of the grayscale image.

This allows us to insert grayscale text, photos or videos between lines AB and CD, and see corresponding spectral images by looking at the line of light at $L_S$ through a dispersive element for analysis ($DE_A$).

To visualize the transformation from grayscale to spectral image, see the Dispersion Diagram in Fig. 4. $DE_S$ disperses the grayscale image in the $x_S$-direction, each of the monochromatic constituents supplying a different image stripe to $L_S$. $DE_A$ rearranges these stripes according to wavelength to form the spectral image.

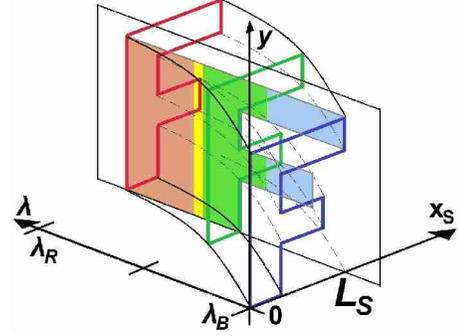

Fig. 4. Imaging principle of Projected-Image Circumlineascopy (PICS).The Dispersion Diagram shows the wavelength distribution of a white, boldface 'F' projected through a dispersive element for synthesis. Being dispersed along the $x_S$-axis, the monochromatic constituents of the grayscale image contribute different image stripes to the $\lambda$-$y$-plane at a given coordinate $x_S$. A dispersive element for analysis ($DE_A$) translates the $\lambda$-$y$-plane into 2 spatial dimensions, thus revealing a rainbow-colored 'F'.

### 3. One setup for both SNS and PICS

An SLM-projector (SONY LCD VPL – CX 70; UHP mercury arc lamp, 2,000 ANSI lumens, 1024 x 768 pixels), connected to a PC, and an amici prism ($\delta_S \approx 4.3°$) as $DE_S$ (unless otherwise specified), were used.

As in Figure 2(a), an image of white lines AB and CD in a black presentation slide was projected at a distance $D_P = $ 2m onto two 0.6m x 0.8m cardboards as S1. With $DE_S$ at $d_S = $ 1.93m, this yielded two spectra, each of width $w_S = $ 0.15m, cf. Figs. 7(a)-(b). The image width was adjusted until Eq. (9) was fulfilled. The purple light at $L_S$ exited the slit between the cardboards.

The light of the calibration spectrum from line AD went at an angle $\theta_i = 0°+/-3°$ through a transmission grating G (grating constant $1/g = 1000$/mm) at a distance $d = $ 3cm behind the slit and onto a 2-D translucent screen S2 at a distance $d_A = $ 27cm behind the grating, as in Figs. 2(b)-(e). From the pixel positions in a digital photograph of diffraction orders m=0 and m=1, $\lambda_{LS} = \lambda_{LS}(y)$ was calculated with the grating formula

$$g(\sin\theta_i + \sin\theta_m) = m\lambda, \quad (9)$$

$\theta_m$ being the diffraction angle.

As a translucent 1-D projection screen for the synthesized light, an uncooked capellini (diameter $w = $ 0.9mm, length $l = $ 26cm) was placed at the slit, and S1 was removed.

For SNS spectral synthesis, grayscale patterns spanned from line AB to CD, cf. Fig. 7(a). The capellini was photographed from the left with a Nikon D80 through a diffraction grating at a distance $d_A = $ 0.4m.

For PICS, grayscale text (cf. Fig. 9(a)), photos (cf. Fig. 8(a), Fig. 11(a)) or videos (played via VLC media player) between lines AB and CD were optimized in contrast and brightness. The capellini was viewed through one or two amici prisms (dispersion angle $\delta_A \approx 4.3°$) or a transmission grating ($1/g = 1000$/mm) as $DE_A$ within a radius $d_A \leq 2$m around the capellini, see Fig. 5.

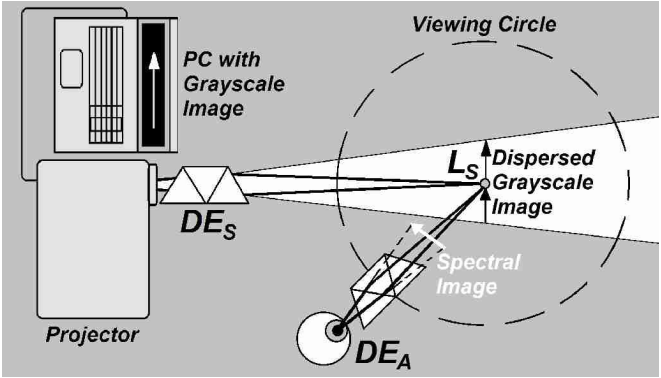

Fig. 5. Setup for SNS, extended for PICS (top view).

## 4. Experimental Results

### A. SNS Spectral Synthesis

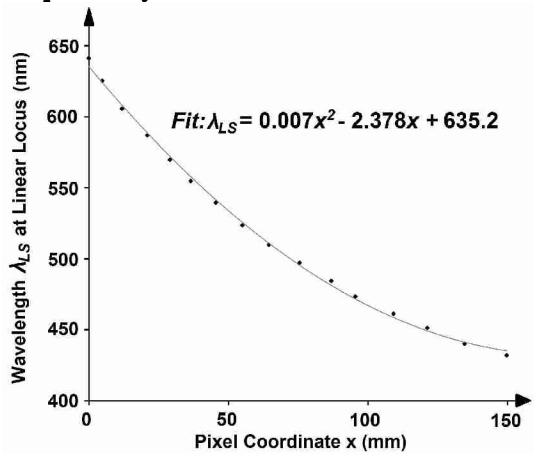

Fig. 6. The fit to the data points constitutes a wavelength calibration.

Peak wavelengths from $\lambda_B = 431$nm to $\lambda_R = 642$ nm were identified and translated into pattern coordinates $x = x(\lambda_{LS}) = 169.9 - [(\lambda_{LS} - 433.2)/0.007]^{0.5}$ using a quadratic fit, see Fig. 6. With a pixel width $\Delta x = 1$mm+/-0.1mm, bandwidths ranged from $\Delta\lambda_{LS}(431\text{nm}) \approx 0.5$nm to $\Delta\lambda_{LS}(642\text{nm}) \approx 5$nm, according to Eq. (6). The grayscale patterns synthesized corresponding spectra, see Fig. 7.

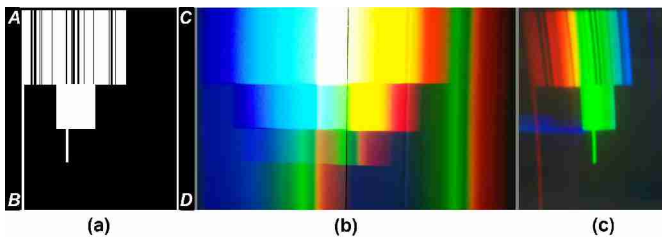

Fig. 7. SNS example spectra. (a) Grayscale pattern. (b) Dispersed graylevel pattern on screen S1. The light passes through a slit where the Newtonian spectra of AB and CD kiss each other. (c) The slit spectra are a rainbow-colored version of the grayscale pattern.

### B. PICS Spectral Images

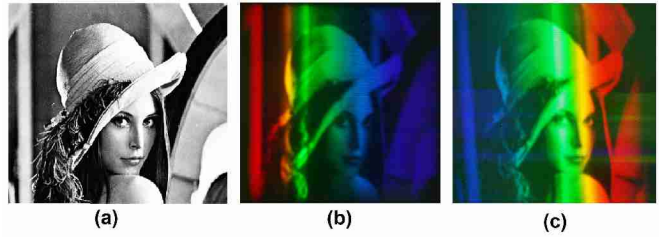

Fig. 8. Example images from PICS. (a) Original grayscale image of 'Lena'. (b) Spectral image of a capellini. (c) High-resolution spectral image of a capellini.

#### 1. Overall appearance

For some views through $DE_A$, see Figs. 8(b)-(c), 9 (b)-(d), and 11 (c)-(d). The virtual spectral images appeared like semitransparent banners attached to the capellini. They turned about the capellini to *face a circumambulating viewer constantly*. While their height looked the same as the height of the light on the capellini, their width increased with the distance $d_A$ from $DE_A$ to the capellini. $DE_A$ with stronger dispersion produced wider images. Their position, size and sharpness did not notably change for a changed distance $d_I$ between viewer and $DE_A$.
If the capellini was viewed obliquely from above or below, the left and right sides of the spectral image were tilted at the same angle as the capellini was due to perspective, see Fig. 11(c). The height was foreshortened; so if the capellini was seen from directly above, the image reduced to a line.

#### 2. Image proportions

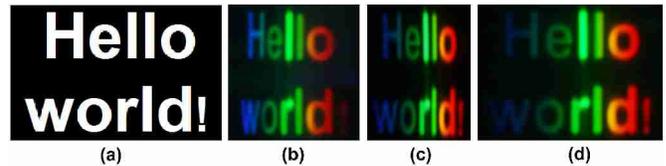

Fig. 9. Different spectral image proportions for different $DE_A$

With an amici prism as $DE_A$, $d_A = 1$m was good for viewing, yet correct image proportions arose at $d_A = 2$m. With the grating as $DE_A$, $d_A = 0.5$m was suitable. At smaller distances, the image was too narrow and blurry; at larger distances, the spectral image was too wide and faint.
The spectral image had different proportions for different $DE_A$. Generally, the width-to-height ratio $w_A/h_A$ of the spectral image was different from the width-to-height ratio $w_S/h_S$ of the grayscale image, see Fig. 9. Specifically, through an amici prism at $d_A = 0.95$m, the text „Hello world!" of Fig. 9(a) had $w_A/h_A = (0.5+/-0.01) w_S/h_S$, see Fig. 9(c). Through a series of amici prisms at distances $d_{A1} = 0.95$m and $d_{A2} = 0.82$m, the text had $w_A/h_A = (0.9+/-0.2) w_S/h_S$, see Fig. 9(d). The letters were evenly spaced in both cases. Through the grating at $d_A = 0.4$m, the text had $w_A/h_A = 0.6(+/-0.01) w_S/h_S$, being squashed toward blue letters, see Fig. 9(b).

*3. Image transformations*

Shifting the capellini to the left of $L_S$ (in the negative $x_S$-direction) removed the right part of the image by shifting the colors. Shifting in the opposite direction effected the opposite.

Shifting the capellini from $L_S$ toward $DE_S$ removed parts from both sides of the image. Shifting the capellini away from $DE_S$ reduced the color range, but left the image intact. In both cases, the image got out of focus, yet did not notably change within a range of about 0.5m. Thus, with a series of capellini within that range, a single grayscale image yielded multiple, almost identical and sharp spectral images at once.

Flipping the orientation of dispersion of $DE_A$ caused the spectral image and its colours to flip, as for $w_4$ in Figs. 12(b)-(c). Flipping the orientation of dispersion of $DE_S$ caused the spectral image to flip geometrically, but the orientation of the color spectrum remained, cf. Fig. 11. Combining both procedures flipped the colors, but not the orientation of the spectral image, cf. Figs. 8(b)-(c).

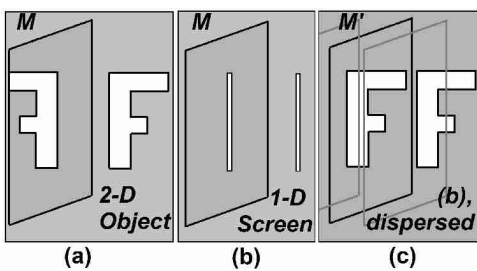

Fig. 10. Mirror Immunity with PICS.(a) A regular 2-D object in front of an upright mirror M is not mirror-immune because its mirror image has reverse orientation.(b) An upright 1-D screen is mirror-immune because its mirror image is the same. (c) Hence, even the 2-D spectral image of the 1-D screen is mirror-immune.

In a mirror parallel to the capellini, *the spectral image was not a mirror image*, but had the same orientation as the spectral image of the capellini before the mirror, as illustrated in Fig. 10.

*4. Multiple-screen, multiple-image PICS*

With grayscale images beside the original one, as in Fig. 11(a), the spectral images of multiple capellini beside $L_S$ were arranged in space as the capellini themselves. They overlapped for some viewing positions, as in Figs. 11(b)-(d). The color result depended on the spatial sequence in which the capellini were viewed. Some capellini produced a spectrally shifted 'doppelgänger' of the spectral image of a neighboring capellini. This happened when the capellini, and the projected grayscale images, were less than a spectrum width $w_S$ apart, as in Fig. 11.

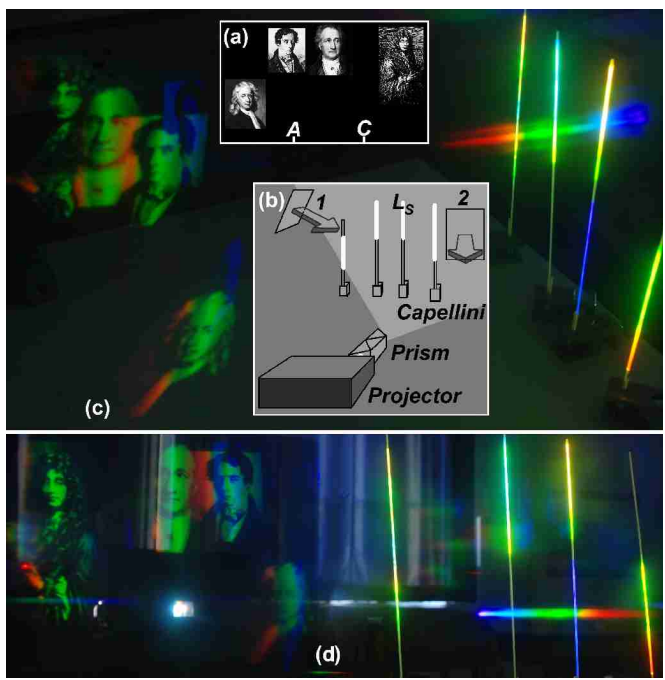

Fig. 11. Multiple-screen, multiple-image PICS. (a) Arrangement of 4 grayscale portraits (Newton, Fresnel, Goethe and Huygens) on a presentation slide. (b) The slide was projected across 4 capellini through a single amici prism with horizontal dispersion as $DE_S$. Photos (c) and (d) were directly taken through a transmission grating at positions 1 and 2. Like rotatable banners attached to the capellini, the spectral portraits always face the viewer. The amici prism at the projector (in photo (d) a white spot) can be flipped horizontally to flip each portrait horizontally.

*5. Image resolution*

Image resolution was enhanced with $L_S$ further from the projector as the grayscale image could be enlarged relative to the presentation slide. Additionally, a second amici prism at the projector allowed the grayscale image to take up almost twice as many pixels horizontally. Thus, horizontal image resolution improved, as in Figs. 8(b)-(c).

Image resolution depended on the orientation of dispersion, apparently because the screen was not perfectly one-dimensional. To investigate how the width $w$ of the projection screen at $L_S$ affects PICS image resolution, a translucent paper was placed at a distance $d_S = 1.5$m from the amici prism to scatter the whole dispersed grayscale image. It was viewed through an amici prism as $DE_A$ from behind the paper while the effective screen width was reduced with a slit aperture directly at the paper (see Fig. 12).

Generally, a $DE_A$ with parallel orientation of dispersion (relative to $DE_S$) made the image even blurrier, while an antiparallel $DE_A$ reduced the blurriness. Specifically, if the spectral images were viewed at a distance $d_A = 1.5$m behind the translucent paper, the difference was extreme: On the one hand, for an antiparallel $DE_A$, the spectral image was sharp for an arbitrarily wide screen, see Fig. 12(b). On the other hand, for a parallel $DE_A$, the spectral image was only sharp for an extremely narrow screen, see Fig. 12(c). The same was true for an equivalent setup with gratings. Synthesized on a single blond human hair of length $l \approx 0.5$m, spectral images were sharpest, yet faint, and movements in a video were easily recognizable.

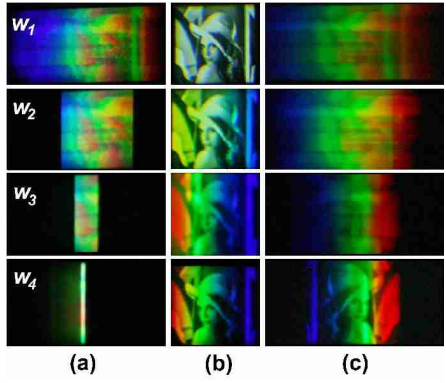

Fig. 12. Screen width and image resolution.(a) The width of a 2-D translucent screen is successively reduced (from $w_1$= 17cm; $w_2$ = 8cm; $w_3$ = 2cm to $w_4$ = 0.5cm) using a slit aperture while the grayscale image of 'Lena' is projected through $DE_S$ from behind. (b) $DE_A$ compensates the dispersion of $DE_S$.(c) $DE_A$ enhances the dispersion of $DE_S$. Uncannily, Lena looks left in the blurry spectral image at $w_1$ and $w_2$, but looks right in the sharp image at $w_4$, appearing Janus-faced at $w_3$.

## 5. Discussion and Analysis

### A. SNS versus other methods of spectral synthesis

Superposing single parts from multiple spectra is as effective in spectral synthesis as modifying a single spectrum, but less efficient: In SNS, only $w/w_S \leq 1\%$ of the dispersed light is used. An equivalent approach has been used to synthesize infrared spectra, but not visible light [10].

The SNS setup in Fig. 2(a) resembles the variable spectrum generator (VSG), but the VSG has a linear variable filter for wavelength selection and a cylindric lens for optical compression [16]. Conveniently, $DE_S$ fulfills both functions.

Wavelength calibration as in Figs. 2(b)-(e) is simple, but presupposes negligible distortion of line AD by $DE_S$, negligible divergence of the dispersed rays from $L_S$, and a distortion-free photograph of the diffraction image. Hence, the error in Fig. 6 was +/-4nm. Although intensity calibration with rectangle ABCD is valid, Eq. (5) requires perfect darkness for 'off' pixels, which is hard to fulfill in practice. Conventional calibration methods are more accurate [2,10].

Whereas the VSG yields bandwidths around 2nm, the SNS light engine can produce much smaller bandwidths thanks to its flexible setup geometry.

As with the VSG, the grayscale pattern intensity may take on 256 values. $I_A(x)$ can be further regulated via the number of 'on' pixels in a given SLM column. This enables the SNS light engine to synthesize spectra with the same quality as the more complex spectral integrator [2].

### B. Properties of PICS Images

*1. Geometry and colors of the virtual spectral image*
How can we quantify the position, size, orientation, and colors of a virtual spectral image?

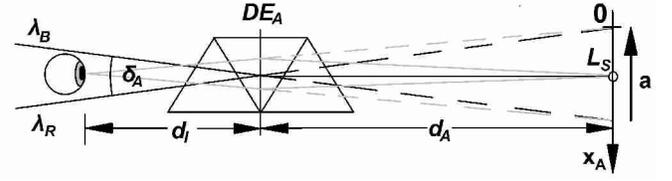

Fig. 13. The virtual spectral image inspected through $DE_A$ (top view) is geometrically analogous to a real image projected through it.

Through an amici prism, the spectral image appears centered at the same position as the line of light itself, see Fig. 13. After all, the prism displaces the monochromatic constituents only in the $x_A$-direction. Through a grating, the spectral image appears to the left and right of the line of light (at m=-1 and m=+1). The spectral image has the same distance to the grating as the line of light, measured from the point where the central ray (for $\lambda_G = 0.5(\lambda_R + \lambda_B)$) to the viewer passes the grating [17]. In this sense, the spectral image is at the same distance $d_A$ from $DE_A$ as the line of light, both for the amici prism, and the grating.

As already Newton found in his Experiment XI in Book I, Part II of his *Opticks* [18], equal but opposite dispersions by $DE_S$ and $DE_A$ reproduce the original image. This allows us to express, analogous to Eq. (1), the dispersive displacement for analysis for a $DE_A$ with dispersion angle $\delta_A$ (see Fig. 13) as
$$a = 2\tan(0.5\,\delta_A)d_A. \qquad (10)$$
While $s$ determines the coloration of image stripes (cf. Fig. 4), $a$ dictates the orientation of the spectral image via the order of the color stripes. Analogous to Eq. (2), $a$ also determines the width of the spectral image
$$w_A = |a| + w, \qquad (11)$$
$w$ being the width of the 1-D screen, which is assumed to be cylindric, here.

Because dispersion in the $x_A$-direction leaves the view unchanged in the y-direction, the height $h_A$ of the virtual spectral image equals the height $h_S$ of the synthesized line of light, or the height of the projected grayscale image:
$$h_A = h_S. \qquad (12)$$

Equations (10)-(12) hold for a prism with negligible magnification, and for a grating. Further, with these equations, we predict the spectral image to have constant position, size, orientation and colours even if $d_I$ varies, as long as $d_A$ is constant. Suppose $DE_A$ is not moved relative to the line of light. Then, for an amici prism, $d_A$ is always constant. In contrast, for a grating, Eq. (9) implies that $d_A$ varies according to the viewer's movements [17]; except along a single moving direction, given by the central ray diffracted to the viewer.

The image transformations described in Section 4 can be understood with Fig. 4: Shifting the 1-D screen along the $x_S$-axis corresponds to shifting the $\lambda$-$y$-plane. Shifting the 1-D screen toward or away from $DE_S$ corresponds to varying the displacement between the monochromatic constituents according to $s$.

*2. Calculating PICS image proportions*
As we saw in the experiment, the spectral image has a different width-to-height ratio than the grayscale image, except at a specified distance $d_A$, depending on $DE_A$.

Let us assume the grayscale image extends from line AB to CD, so $w_S/(|s|+\Delta x) = 1$, as in Eq. (2). Then, based on Eqs. (11)-(12), the width-to-height ratio $w_A/h_A$ of the spectral image relates to that $w_S/h_S$ of the grayscale image as follows:

$$\frac{w_A}{h_A} = \left[\frac{|\tan(0.5\delta_A)d_A|+0.5w}{|\tan(0.5\delta_S)d_S|+0.5\Delta x}\right]\frac{w_S}{h_S} . \quad (13)$$

Thus, to obtain a spectral image with a desired width-to-height ratio, we may adapt the height $h_S$ of the grayscale image, use different dispersive elements, or change their distances to the projection screen, according to Eq. (13).

To compensate distortions *within* the spectral image, we adapt the proportions within the grayscale image according to the wavelength scales $x=x(\lambda_{LS})$ of $DE_S$ and $x_A=x_A(\lambda_{LS})$ of $DE_A$, as Fig. 4 suggests.

### 3. Calculating PICS image resolution

To grasp PICS image resolution intuitively, cut a picture into narrow vertical stripes of width $w$ and stack them. This is analogous to Fig. 12(c) at $w=w_4$. If you then spread the stripes out in the original direction, the original image appears; analogous to Fig. 12(b) at $w_4$. If, instead, you spread the stripes out in the opposite direction, a mirror image, less sharp, emerges; analogous to Fig. 12(c) at $w_4$. This analogy also explains why 'Lena' turns her face in Fig. 12(c) as the wide screen is narrowed.

To describe PICS image resolution quantitatively, we refer to Dispersion Diagrams [15], as in Fig. 14. Let us discuss the straightforward case of amici prisms as $DE_A$ and $DE_S$. As shown in Figs. 14(a) and (d), $DE_S$ shears the wavelength distribution of the grayscale image according to dispersion vector **s**. Thus, at $L_S$, all grayscale pixels translate into spectral pixels. These are superposed in a spectral stack of width $w \leq \Delta x$. Now, consider a viewer who looks in the direction of light projection. Then, the unit vector $\hat{x}_A$ - which always points to the right of the viewer (see Fig. 13) - is parallel to $\hat{x}_S$. $DE_A$ shears the spectral stack, including the spectral pixels, according to dispersion vector

$$\boldsymbol{a} = a\hat{x}_A . \quad (14)$$

From Figs. 14(b)-(c) and (e)-(f), we geometrically derive the pixel width of the spectral image:

$$\Delta x_A = \frac{|a|}{N-1} + \frac{|a+s|}{N-1}\left(\frac{w}{\Delta x}\right) . \quad (15)$$

With $|s|=(N-1)\Delta x$, we obtain the spectral pixel width

$$\Delta x_A = \left|\frac{a}{s}\right|\Delta x + \left|\frac{a}{s}+1\right|w . \quad (16)$$

Note in Fig. 14 that spectral pixels partly overlap, depending on **a**. Two spectral pixels are just resolved when their distance equals their FWHM, labeled $\Delta x_H$. From Fig. 14, we obtain, depending on the relative dispersive displacement for analysis a/s,

$$\Delta x_H = \left|\frac{a}{s}\right|\Delta x , \text{ for } \frac{a}{s} \leq -0.5 \quad (17)$$

and $\Delta x_H = \Delta x_A - \left|\frac{a}{s}\right|w$, for $\frac{a}{s} \geq -0.5$ . (18)

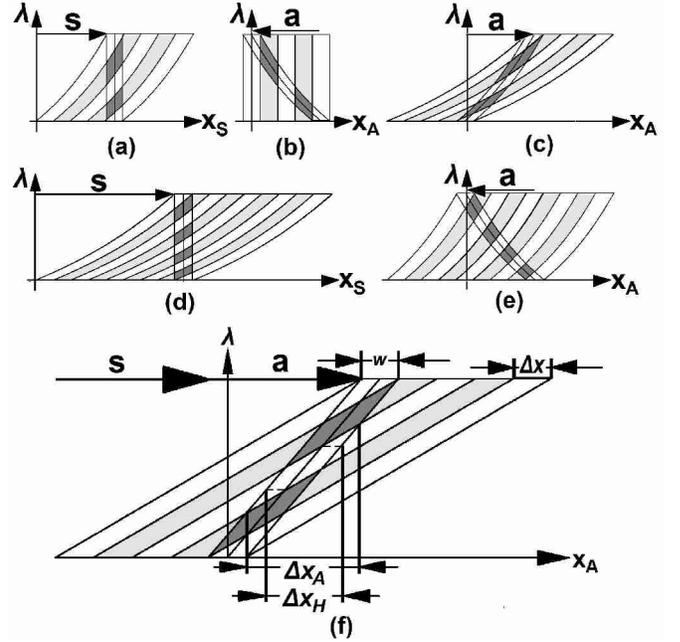

Fig. 14. Deriving PICS spectral image resolution from Dispersion Diagrams. In (a) and (d), the wavelength distribution of a grayscale image, dispersed by $DE_S$ according to dispersion vector **s**, contains a spectral stack of width w at the linear locus (accentuated) that represents all grayscale pixels as spectral pixels. In (b), (c) and (e), additional dispersion by $DE_A$ shears the spectral stack according to dispersion vector **a**, the spread-out spectral pixels forming a spectral image whose image resolution depends on the relative dispersive displacement for analysis, $a/s$.

(a) At $a/s = 0$, all spectral pixels are superposed. (b) At $a/s = -1$, all spectral pixels lie next to each other. (c) At $a/s = +1$, spectral pixels partially overlap. In (d), where $a/s = 0$, and in (e), where $a/s = -0.5$, the dispersion of $DE_S$ is twice as strong as in (a)-(c), allowing more pixels to be represented at the linear locus, yielding higher pixel resolution compared to (b). (f) The spectral pixel width $\Delta x_A$, and the FWHM of a spectral pixel, $\Delta x_H$, depend on the width w of the line of light at the linear locus, on the grayscale pixel width $\Delta x$, and the relative dispersive displacement for analysis, $a/s$.

Finally, let us define spectral image resolution as the number of spectral pixels that would - based on our resolution criterion - fit within the width of the spectral image, namely as

$$R_A = \frac{w_A}{\Delta x_H} . \quad (19)$$

This is in analogy to the grayscale image resolution, $R_S = w_S/\Delta x$. The asymmetry in the graphs in Fig. 15 depicts our observation that image resolution depends on the orientations of dispersion of $DE_S$ and $DE_A$. Only for vanishing $w/\Delta x$, which means a perfectly 1-D line of light, the asymmetry vanishes.

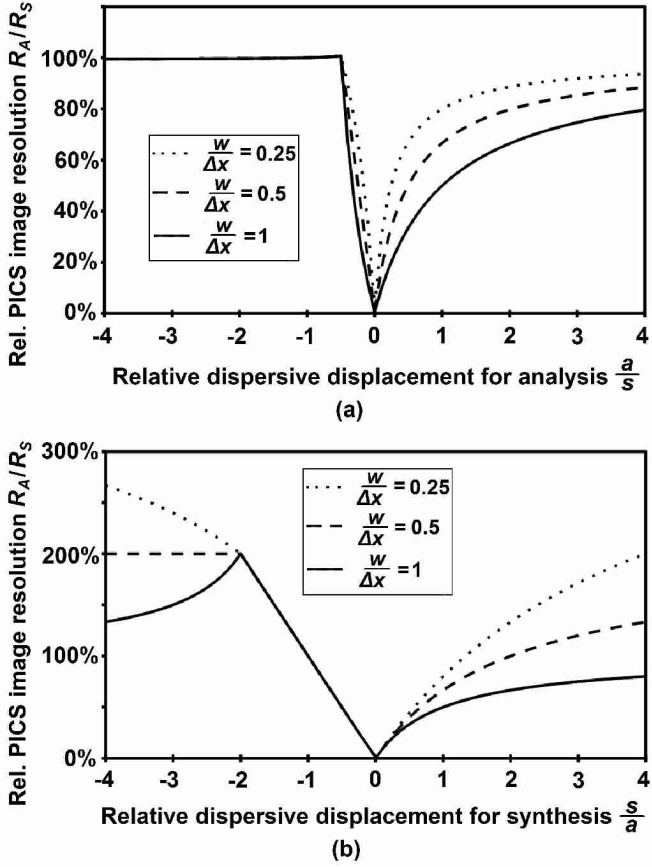

Fig. 15. PICS spectral image resolution in terms of grayscale image resolution, calculated with Eqs. (16)-(18). (a) Image resolution decreases by decreasing the dispersive displacement for analysis $a$, at fixed dispersive displacement for synthesis $s = 150\Delta x$. (b) Image resolution can be improved by increasing the dispersive displacement for synthesis $s$, at fixed dispersive displacement for analysis $a = 150\Delta x$.

Now, consider an *arbitrary* azimuthal viewing direction. Suppose the translucent screen at $L_S$ be flat instead of cylindric, and contain $\hat{x}_S$. Then, dispersion vector **a** does not shear the spectral stack itself, but its *orthogonal projection* onto the $x_A$-axis. Accordingly, we introduce $s_A = \mathbf{s}\,\hat{x}_A$. Generalizing Eq. (16), we get

$$\Delta x_A = \left|\frac{a}{s}\right| \Delta x + \frac{|a + s_A|}{|s|} w \,. \quad (20)$$

Generalizing Eqs. (16) and (17), we obtain

$$\Delta x_H = \left|\frac{a}{s}\right| \Delta x \,, \text{ for } \frac{a}{s_A} \le -0.5 \quad (21)$$

and $\quad \Delta x_H = (\Delta x - w)\left|\frac{a}{s}\right| + w\left|\frac{a+s_A}{s}\right| \,, \text{ for } \frac{a}{s_A} \ge -0.5 \,. \quad (22)$

*4. Suggested applications of PICS*

*360°-visibility* offers applications from text display to image projection to object tracking [19]. Imagine a cinema where viewers with diffraction glasses sit around a translucent 1-D screen. (A reflective metal rod of diameter $w = 2$mm offers ca. 310°-visibility, as I found in preliminary experiments.) A grating constant that is inversely proportional to $d_A$ enables correct image proportions. Alternatively, a cylindric transmission grating around the 1-D screen could display correctly-proportioned spectral images to anyone around it.

Being semitransparent, virtual images may be superposed onto an object or image, whether for geometric comparison or color experiments. Further, three virtual spectral images of a single 1-D screen may compose a real-color image. To obtain the RGB image components, their grayscale versions are inserted in the corresponding intervals (600-700nm, 500-600nm, or 400nm-500nm, respectively; cf. [20]) between AB and CD. However, their superposition requires a specially designed $DE_A$.

The 1-D screen takes up little space and material. Besides, its light does not disturb a disinterested individual. Moreover, the relevant light beam is narrow, allowing projection in confined or crowded spaces, even for large images. This solves the problem stated in a recently published paper [21].

Being metameric, the synthesized line of light does not betray the image to viewers without or beyond $DE_A$. This is valuabe in police interrogation, medical communication, advertising, and beyond.

*Mirror immunity* is intrinsic to a 1-D screen. Spectral images may be multiplied or delivered elsewhere via mirrors parallel to the 1-D screen, without ever changing image orientation.

### C. Unifying spectral synthesis and spectral encoding

For PICS, 2-D images are spectrally encoded in 1D. Spectral encoding, whereby locations are translated into wavelengths of light, has already been applied to 2-D image acquisition via 0-D or 1-D apertures, whether in spectrally-encoded endoscopy (SEE) [22], wavelength-multiplexed microscopy [23], or modern pseudoscopy [24]. Spectral encoding was also proposed for the transmission of a 2-D image via an optical fiber [25-27]. Still, it has not yet been applied to video projection.

Until now, SLM-based light engines were thought to encode wavelengths as patterns on the SLM. Looking back on SNS and PICS, we may now state the reverse: SLM-based light engines encode images as wavelengths. This makes any of these light engines suitable for PICS.

A precursor to SNS and PICS is Newton's Experiment I in Book I, Part II of his *Opticks* [18]. Focused on proving his theory, he did not see the practical value of the experiment, however.

### 6. Conclusion

By projecting a grayscale image through a prism or grating, we obtain light at a linear locus $L_S$ whose spectrum is a rainbow-colored version of the grayscale image.

This provides i) a method for synthesizing light with a desired spectral power distribution, called Superposition of Newtonian Spectra (SNS); and ii) a method for viewing 2-D images that are spectrally encoded on a 1-D projection screen, called projected-image circumlineascopy (PICS).

For SNS, an SLM-projector and a prism or grating were used. This setup is the simplest among equally effective SLM-based tunable light sources; the trade-off being considerable light loss.

For PICS, grayscale text, photos and videos from an SLM-projector were dispersed by a prism across an upright capellini. If the capellini was viewed through another prism or grating, rainbow-colored versions of the grayscale images appeared. Floating in midair, the semitransparent images were correctly oriented for any azimuthal viewing angle, even if the capellini was reflected in an upright mirror. Real-color, mirror-immune, surround-view images are achievable with PICS by superposing three virtual spectral images of a single 1-D screen, yet a three-component viewing device needs to be designed for the RGB mixture. An advanced version of PICS is conceivable where a 0-D point of light is dispersed into 2 dimensions, yielding a 2-D spectral image that has correct image proportions and constant apparent size, for *any* viewing position.

Geometric optics were used in tandem with Dispersion Diagrams to visualize the transformation from grayscale to spectral image, and to derive formulas for SNS spectral resolution, PICS image proportions, and PICS image resolution.

Further research could investigate PICS image transformations for variations of the setup, for example with a diagonal 1-D screen. Beyond that, a systematic treatment of generic virtual images in spatial relation to the viewer would be useful.

Besides the SNS tunable source, all SLM-based light engines were shown to project wavelength-encoded images. These are not limited to abstract patterns, but may as well be concrete images.

### Acknowledgments

I thank Prof. Dr. Florian Theilmann for starting our exploration into spectra in 2010, for commenting on some drafts, and for sending me an amici prism for the experiments.